# Experimental tests of the full spin torque conductivity tensor in epitaxial IrO$_2$ thin films

Michael Patton[1], Daniel A. Pharis[3], Gautam Gurung[9,10], Xiaoxi Huang[3], Gahee Noh[5], Evgeny Y. Tsymbal[4], Si-Young Choi[5,6,7], Daniel C. Ralph[3,8], Mark S. Rzchowski[2], Chang-Beom Eom[1*]

[1]Department of Materials Science and Engineering, University of Wisconsin-Madison, Madison, Wisconsin 53706, United States.

[2]Department of Physics, University of Wisconsin-Madison, Madison, Wisconsin 53706, United States.

[3]Cornell University, Ithaca, New York 14853, United States.

[4]Department of Physics and Astronomy & Nebraska Center for Materials and Nanoscience, University of Nebraska, Lincoln, NE 68588, United States.

[5]Department of Materials Science and Engineering, Pohang University of Science and Technology, Pohang, Gyeongbuk 37673, Korea.

[6]Center for Van der Waals Quantum Solids, Institute for Basic Science (IBS), Pohang 37673, Republic of Korea

[7]Semiconductor Engineering, Pohang University of Science and Technology (POSTECH), Pohang, 37673, Republic of Korea

[8]Kavli Institute at Cornell for Nanoscale Science, Ithaca, New York 14853, United States.

[9]Clarendon Laboratory, Department of Physics, University of Oxford, Parks Road, Oxford, OX1 3PU UK

[10]Trinity College, University of Oxford, Oxford, OX1 3BH UK

*Corresponding author: Chang-Beom Eom

Email: eom@engr.wisc.edu

**This PDF file includes:**

    Main Text
    Figures 1 to 3

**Abstract**

Unconventional spin-orbit torques arising from electric-field-generated spin currents in anisotropic materials have promising potential for spintronic applications, including for perpendicular magnetic switching in high-density memory applications. Here we determine all the independent elements of the spin torque conductivity tensor allowed by bulk crystal symmetries for the tetragonal conductor $IrO_2$, via measurements of conventional (in plane) anti-damping torques for $IrO_2$ thin films in the high-symmetry (001) and (100) orientations. We then test that rotational transformations of this same tensor can predict both the conventional and unconventional anti-damping torques for $IrO_2$ thin films in the lower-symmetry (101), (110), and (111) orientations, finding good agreement. The results confirm that spin-orbit torques from all these orientations are consistent with the bulk symmetries of $IrO_2$, and show how simple measurements of conventional torques from high-symmetry orientations of anisotropic thin films can provide an accurate prediction of the unconventional torques from lower-symmetry orientations.

**Significance Statement**

Unconventional spin transport offers advantages such as control over the spin polarization direction in low power memory devices based on anisotropic materials. Here, we study spin Hall conductivity in epitaxial $IrO_2$ thin films, and for the first time show that symmetry transformations between different epitaxial orientations match experimental measurements of these properties. This establishes that spin-orbit torques in a particular orientation can accurately predict the unconventional torques for lower symmetry orientations. It addresses previous comparisons of experimental and theoretical results, and offers a new approach towards building epitaxial materials with designed large unconventional spin transport.

**Main Text**

**Introduction**

The linear-response spin current generated by an applied electric field within a material is described by a third-rank spin Hall conductivity (SHC) tensor $\sigma_{ij}^s$, where *s* is the index for the spin polarization direction, *i* for the spin flow direction, and *j* for the applied electric field direction. If the spin current is absorbed by an adjacent magnetic layer, it will apply a torque that can efficiently reorient the magnetization. The 27 elements within $\sigma_{ij}^s$ are often highly constrained by symmetry, and in commonly-used high-symmetry materials most of the elements are zero because symmetries can require the spin polarization, spin flow, and electric field to be mutually orthogonal. We will term the torques generated by such a spin current as "conventional". For some applications, however, unconventional spin torques are highly desired. In particular, out-of-plane anti-damping torques associated with tensor elements of the form $\sigma_{zj}^z$ (where *z* is the direction normal to the device plane) are needed to drive efficient anti-damping switching of magnetic memory devices with perpendicular magnetic anisotropy. Such out-of-plane anti-damping torques have been realized using spin-source materials in which the symmetry constraints are relaxed by very low crystal symmetries or magnetic ordering[1–12] or by interfacial effects[13–15]. We have also recently proposed a simple alternative strategy for generating out-of-plane anti-damping torques -- to use a relatively high-symmetry but anisotropic material as the spin source (e.g., a tetragonal or orthorhombic structure) and to grow thin films with a growth axis tilted away from any high symmetry direction (e.g., tilted in a (101) or (111) orientation). In this case, the tilt of the crystal axes relative to the sample plane can break the necessary symmetries to allow a nonzero value for $\sigma_{zj}^z$ and an associated unconventional torque. We demonstrated this qualitatively in ref. [12] for the tetragonal material $IrO_2$. Here we test this approach quantitatively. By measuring the electric-field generated spin torque for (001) and (100) thin films of $IrO_2$ we determine all of the independent elements of the spin torque conductivity tensor associated with the absorbed spin current, and then we test whether rotational transformations of this single tensor can provide consistent quantitative predictions for the torques in three lower-symmetry film orientations: (110), (101), and (111). We find excellent agreement between the measurements and the predictions of the rotated tensor for both conventional and unconventional torques in all 5 crystal orientations.

**Results and Discussion**

First, one note regarding notation. The electric-field-generated torque applied to the magnetic layer in a spin-source/magnet bilayer will differ from the spin current generated in the spin-source layer by an interfacial transmission coefficient which describes what fraction of the spin current is absorbed by the magnetic layer. Spin-torque experiments therefore do not measure the spin current directly. In the following, we will assume that the interfacial transmission coefficient is to a good approximation a constant (*T < 1*) that does not depend on the thin-film orientation or the spin orientation, so that for purposes of calculating torques we can define a spin torque conductivity (STC) tensor as simply proportional to the spin Hall conductivity tensor: $\tau_{ij}^s = T\sigma_{ij}^s$. The purpose of this paper is to test this assumption, i.e., to test the degree to which rotational transformations of a single STC tensor can give an accurate description of the measured torques for different crystal orientations.

The bulk crystal symmetries for the tetragonal structure of rutile $IrO_2$ dictate that the spin Hall conductivity tensor can be defined in terms of three independent elements. For the corresponding STC tensor we will call these elements *a*, *b*, and *c*. If we define basis vectors in terms of the $IrO_2$ crystal axes as X = [100], Y = [010], and Z = [001], the most general form allowed for the STC tensor is

$$\tau^X \quad\quad\quad \tau^Y \quad\quad\quad \tau^Z$$

$$\begin{bmatrix} 0 & 0 & 0 \\ 0 & 0 & b \\ 0 & -a & 0 \end{bmatrix} \quad \begin{bmatrix} 0 & 0 & -b \\ 0 & 0 & 0 \\ a & 0 & 0 \end{bmatrix} \quad \begin{bmatrix} 0 & c & 0 \\ -c & 0 & 0 \\ 0 & 0 & 0 \end{bmatrix}.$$

Examples of spin currents corresponding to purely the *a*, *b*, and *c* processes are depicted in Fig. 1c. If a thin film of $IrO_2$ is grown in an orientation different than (001), one can perform a change of basis in order to define the STC tensor relative to basis vectors in the plane and perpendicular to the thin film. This is achieved through a rotational transformation of the form:

$$\tau^s_{ij} = \sum_{l,m,n} R_{il} R_{mj} R_{sn} \tau^n_{lm} \tag{1}$$

where $R_{il}$ are the elements of the appropriate rotation matrix. For example, for a (100) oriented $IrO_2$ film, using the basis vectors X = [010], Y = [001], Z = [100], the STC tensor takes the form

$$\tau^X \quad\quad\quad \tau^Y \quad\quad\quad \tau^Z$$

$$\begin{bmatrix} 0 & 0 & 0 \\ 0 & 0 & a \\ 0 & -b & 0 \end{bmatrix} \quad \begin{bmatrix} 0 & 0 & -c \\ 0 & 0 & 0 \\ c & 0 & 0 \end{bmatrix} \quad \begin{bmatrix} 0 & b & 0 \\ -a & 0 & 0 \\ 0 & 0 & 0 \end{bmatrix}.$$

The transformed forms of the STC tensor appropriate for the (101), (110), and (111) thin-film orientations of $IrO_2$ are listed in the supplemental information.

To determine the full STC tensor in $IrO_2$, the three elements *a, b,* and *c* must be experimentally measured. Spin-torque measurements in spin-source/magnet bilayer samples are only sensitive to spin currents flowing perpendicular to the sample plane, because only for this flow direction can the spin current be transmitted to the magnetic layer to exert a torque. This means that only the elements in the bottom rows (i.e., $\sigma^S_{Zj}$) of the STC tensor are accessible. The first element, *a*, can be measured using the (001)-orientated film, where an electric field is applied along the [010] direction resulting in a [$\bar{1}$00]-polarized spin current flowing in the out-of-plane [001] direction as indicated in Fig. 1b. The second element, *b*, can be measured using the (100) orientation, with an electric field applied along the in-plane [001] direction resulting in a [0$\bar{1}$0]-polarized spin current flowing in the out-of-plane [100] direction seen in Fig. 1e. The third and last term, *c*, can also be measured in the (100) orientation. An electric field applied along the [010] direction results in a [001]-polarized spin current flowing in the out-of-plane [100] direction seen in Fig. 1g.

To measure the electric-field-generated torques experimentally, high crystalline thin films of $IrO_2$ were grown via RF magnetron sputtering on different orientations of $TiO_2$ substrates, then capped *in situ* with ferromagnetic permalloy (Py), and patterned into device structures using the same methods described in our previous report[12]. High resolution x-ray diffraction (HRXRD) demonstrated a single high crystalline phase for all five thin-film IrO$_2$ orientations studied in this paper: (001), (100), (110), (101), and (111) (see Fig. 2a-e), and scanning transmission electron microscopy (STEM) also demonstrated sharp IrO$_2$/Py interfaces using STEM as shown in Fig. 2f-i. The spin-torque ferromagnetic resonance (ST-FMR) technique was used to characterize the SHC components by measuring ST-FMR resonance spectra as a function of sweeping the magnetic-field magnitude for a series of different directions of the magnetic-field within the plane of the sample. (See Methods for more details.)

The symmetric and antisymmetric ST-FMR amplitudes for the (001) and (100) orientations are shown in Fig. 1a,d, and f as a function of the angle of an in-plane applied magnetic field. The symmetric signals allow us to determine the three independent elements that define the anti-damping STC tensor of IrO$_2$ (see Methods). The first term $a$, determined in the (001) orientation, has a value of 520 ± 19 ($\frac{\hbar}{e}$ ($\Omega$ cm)$^{-1}$). The second term $b$, determined in the (100) orientation with an electric field applied along the [001] direction, has a value of 238 ± 5 ($\frac{\hbar}{e}$ ($\Omega$ cm)$^{-1}$). The third term $c$ determined in the (100) orientation with electric field along the [010] direction, has a value of 493 ± 15 ($\frac{\hbar}{e}$ ($\Omega$ cm)$^{-1}$). As required by symmetry for the (001) and (100) orientations with these electric-field directions, we detect no unconventional torque components.

To test whether this same tensor gives a quantitative description of both the conventional and unconventional torques for other thin-film orientations, we also performed ST-FMR measurements for (110), (101), and (111) IrO$_2$ thin films. For each thin-film orientation, the measurements were made using various directions ($\psi$) of in-plane electric field (and hence charge current), with $\psi$ measured relative to the direction defined as the X axis for that orientation. For convenience in distinguishing the conventional and unconventional torques, we will plot the measured torque tensor components using a different set of coordinate axes (using lower-case x, y, and z), in which the x axis is along the applied electric field direction and z remains normal to the sample plane. The anti-damping torque components we measure for each orientation are then the conventional in-plane torque perpendicular to $E$ (the tensor component $\tau_{zx}^y$), the unconventional in-plane torque parallel to E ($\tau_{zx}^x$), and the unconventional out-of-plane torque ($\tau_{zx}^z$). The results are shown as the symbols within the Fig. 3 graphs. The solid lines in Fig. 3 are the predicted STC values from the rotated experimental tensor, using the values of $a$, $b$, and $c$ as determined above with no adjustable fit parameters. We find that, by using the STC tensor elements from the experimental (001) and (100) ST-FMR results and assuming the torque is governed by the bulk symmetries of IrO$_2$, the tensor rotation gives a good description of both the conventional and unconventional torque components for the other orientations. (Supplemental Note 3 shows zoomed-in plots of the unconventional torques.) For the (100), (110), (101), and (111) orientations, an unconventional in-plane STC was observed and followed the expected sin(2$\psi$) angular dependence, with a magnitude within 30 ($\frac{\hbar}{e}$ ($\Omega$ cm)$^{-1}$) of the value predicted by the tensor rotation.. For the (101) and (111) orientations, unconventional out-of-plane anti-damping torque was also present, following an expected sin($\psi$) dependence, with a magnitude within 10 ($\frac{\hbar}{e}$ ($\Omega$ cm)$^{-1}$) compared to the expectation from the tensor rotation. Both types of unconventional torques are significantly weaker than the conventional in-plane anti-damping torques, but this is fully consistent with the tensor rotation given the measured values of $a$, $b$, and $c$. For example, in the (111) orientation, the largest amplitude predicted by the tensor rotation for the out-of-plane anti-damping torque is predicted to be $(a - c)/2$ = 14 ± 24 ($\frac{\hbar}{e}$ ($\Omega$ cm)$^{-1}$) (see supplemental information), or less than 3% of the conventional torque for the (001) orientation. For the conventional in-plane anti-damping torque, the tensor rotation correctly predicts the observed dependence on the electric-field angle, but the magnitudes in some cases show somewhat larger deviations than for the unconventional torques. For example, the predicted conventional in-plane torque for the (111) orientation is about 20% lower than the measurements. One possible explanation for this could be due to the surface quality of the (001) and (100) orientations compared to (111). Due to the low surface energy of the (111) orientation, the interface between IrO$_2$ and Py is much sharper compared to the (001) and (100) orientations which could change the spin transparency at the interface. Compared to previous measurements for IrO$_2$ (001) and (110) grown by reactive oxide molecular beam epitaxy, the torque magnitudes for our (110) samples are in close agreement, whereas the torques for (001) orientation in our films are about a factor of two smaller.[16] This difference could be due to an Ir spacer layer included in the previous work or due to different growth techniques. The results in Fig. 3 correspond to anti-damping torque components. In addition, unconventional field-like torques were present in the (100), (110), (101), and (111) orientations which we attribute to anisotropic resistances within these orientations (see Fig. S9 in Supplemental Note 3)[4,17].

Despite the consistency of the experimentally-determined anti-damping spin-torque tensor for different crystal orientations, the experimental results are inconsistent with density-functional-theory (DFT) calculations of the spin Hall conductivity. The DFT predictions for the elements of the spin Hall conductivity tensor corresponding to the *a*, *b*, *c* parameters of the STC tensor are $a_{SHC}$ = 254 ($\frac{\hbar}{e}$ ($\Omega$ cm)$^{-1}$), $b_{SHC}$ = 162 ($\frac{\hbar}{e}$ ($\Omega$ cm)$^{-1}$), and $c_{SHC}$ = 18 ($\frac{\hbar}{e}$ ($\Omega$ cm)$^{-1}$). Despite the fact that the SHC magnitudes should be larger than the STC magnitudes on account of the interfacial spin transmission factor ($T$ < 1), the spin Hall conductivities predict values that are too small, by more than a factor of 2 for the *a* parameter and by a factor of 27 for the *c* parameter. This leads to the measured conventional spin-orbit torques being significantly larger than expected from the DFT predictions. In contrast, the measured unconventional out-of-plane torque for low-symmetry crystal orientations is nevertheless much smaller than predicted by DFT. This is because the out-of-plane anti-damping torque is proportional to $a - c$, and this difference is much smaller for the torque parameters than the difference predicted by DFT, $a_{SHC} - c_{SHC}$. The poor agreement between measurements of spin-orbit torque and DFT predictions is true not only for IrO$_2$, but also for most materials including the prototypical spin source Pt [18,19]. This indicates that essential physics is still missing from this comparison. Nevertheless, since our tensor-rotation analysis depends only on the bulk symmetries of IrO$_2$ with no assumptions about microscopic mechanisms, predictions based on the rotated-tensor analysis remain valid and accurate.

**Conclusion**

We have experimentally determined the full anti-damping spin torque conductivity tensor for IrO$_2$, and showed that this single tensor provides consistent and accurate results of the measured electric-field-driven torques for five different thin-film orientations, including both conventional and unconventional torques. The good agreement between the experimental measurements of the anti-damping spin torques and the predictions from the tensor rotation in Fig. 3 confirm that the electric-field-induced torques generated by IrO$_2$ originate from bulk spin currents within the IrO$_2$, with perhaps a minor contribution to the conventional spin torque in the (111) orientation from an additional interfacial effect. We observe no indication of large differences in interfacial spin transmission for different crystal orientations that would invalidate the tensor analysis. In addition, the tensor analysis show that it is possible to fully characterize the spin-torque tensor of an anisotropic material using measurements of conventional spin-orbit torque for selected high-symmetry crystal orientations, and then to obtain accurate predictions of the unconventional torques for lower-symmetry crystal orientations by means of a simple tensor rotation.

**Materials and Methods**

*Sample growth, fabrication, and characterization.*

Epitaxial IrO$_2$ was grown on TiO$_2$ (001), (100), (110), (101) and (111) substrates by RF magnetron sputtering followed by *in situ* growth of ferromagnetic permalloy Ni$_{81}$Fe$_{19}$ (Py). The IrO$_2$ films were grown at 400°C at a pressure of 30 mTorr with 10% oxygen partial pressure. The target power was 20 W. After growth the sample was cooled in an O$_2$ atmosphere. Py was then grown in situ at room temperature, 4 mTorr of Ar, power of 35W, and a background pressure of 3E-7 Torr. The samples were then fabricated using photolithography and ion beam milling, followed by sputter deposition of 100 nm Pt/10nm Ti and lift off techniques for the electrodes.

*ST-FMR measurements*

During the ST-FMR measurements, a microwave current was applied at a fixed frequency (5-12 GHz) and fixed power (10-13 dBm) while sweeping an in-plane magnetic field through the Py resonance conditions from 0 to 0.15 T. The microwave current was modulated at a fixed frequency of 437 Hz and the mixing voltage across the device was measured using a lock-in amplifier. The mixing voltage was fitted vs applied field to extract the symmetric and antisymmetric Lorentzian components. For the angular-dependent ST-FMR, the applied field was rotated in-plane 360° and the symmetric and antisymmetric components were plotted as a function of angle. The out-of-plane (T$_\perp$) and the in-plane (T$_\parallel$) torques are proportional to the mixing voltage V$_{mix}$ as the ferromagnetic layer goes through its resonance condition, which can be fitted as a sum of a symmetric and an antisymmetric Lorentzians:

$$V_{mix,S} = -\frac{I_{rf}}{2}\left(\frac{dR}{d\varphi}\right)\frac{1}{\alpha(2\mu_0 H_{FMR}+\mu_0 M_{eff})}\text{T}_\parallel$$

$$V_{mix,A} = -\frac{I_{rf}}{2}\left(\frac{dR}{d\varphi}\right)\frac{\sqrt{1+M_{eff}/H_{FMR}}}{\alpha(2\mu_0 H_{FMR}+\mu_0 M_{eff})}\text{T}_\perp \ .$$

Here $I_{rf}$ is the RF current calibrated using Joule heating experiments, *R* is the resistance of the device, $\varphi$ is the magnetization angle with respect to the applied current, $\alpha$ is the Gilbert damping coefficient determined from the ST-FMR linewidth (giving values in the range $\alpha$ = 0.013 - 0.037, $\mu_0 H_{FMR}$ is the resonance field, and $\mu_0 M_{eff}$ is the effective magnetization ranging from 0.6 - 0.85 T. The effective magnetization of Py is be obtained using Kittel's equation $f =$ $\frac{\gamma}{2\pi}\sqrt{(H_{FMR}+H_K)(H_{FMR}+H_K+M_{eff})}$ where $\gamma$ is the gyromagnetic ratio and $H_K$ is the in-plane anisotropy field. Unconventional torques are characterized based on the dependence of the measured torques on the angle $\varphi$ of the magnetic field:

$$T_{\parallel} = T_{x,AD}\sin(\varphi) + T_{y,AD}\cos(\varphi) + T_{z,FL}$$

$$T_{\perp} = T_{x,FL}\sin(\varphi) + T_{y,FL}\cos(\varphi) + T_{z,AD}$$

$V_{mix,S}$ and $V_{mix,A}$ can then be expressed in the form of $\sin(2\varphi)(T_{x,AD}\sin(\varphi) + T_{y,AD}\cos(\varphi) + T_{z,FL})$ and $\sin(2\varphi)(T_{x,FL}\sin(\varphi) + T_{y,FL}\cos(\varphi) + T_{z,AD})$, respectively. Here the subscript x denotes a component associated with the spin polarization direction aligned with the applied electric field, and z the component associated with spin polarization normal to the sample plane. Additionally, here capital T represents the torque on the magnetic moment which is different than lowercase τ used denote the spin torque conductivity in the main text. The spin torque conductivity $\tau^i_{jk}$ can then be determined using $\tau^i_{jk} = \frac{\theta_i}{\rho_{IrO2}}\frac{\hbar}{2e}$, where $\rho_{IrO2}$ is the resistivity and $\theta_i$ is defined as:

$$\theta_i = T_{i,AD}\frac{2e\mu_0 M_s t_{FM}}{\gamma \hbar J}$$

where $t_{FM}$ is the thickness of the ferromagnetic layer (Py or Ni) and $J$ is the charge current density in the IrO$_2$. The resistivities were determined using Van der Pauw measurements and found to be 105 $\mu\Omega cm$ for (001), 226 $\mu\Omega cm$ for (100) along the [010] direction, 160 $\mu\Omega cm$ for (100) along the [001] direction, 100 $\mu\Omega cm$ for (101) along the [010] direction, 150 $\mu\Omega cm$ for (101) along the [10-1] direction, 95 $\mu\Omega cm$ for (111) along the [11-2] direction, and 105 $\mu\Omega cm$ for (111) along the [1-10] direction.

*Theoretical Calculations*

DFT calculations were performed using a Quantum-ESPRESSO code[20]. The plane-wave pseudopotential method with the fully relativistic ultrasoft pseudopotentials[21] was employed in the calculations. The exchange and correlation effects were treated within the generalized gradient approximation (GGA)[22]. The plane-wave cut-off energy of 40 Ry and a 16 × 16 × 16 k-point mesh in the irreducible Brillouin zone were used in the calculations. Spin-orbit coupling was included in all the calculations.

The spin Hall effect is given by:

$$\sigma^k_{ij} = \frac{e^2}{\hbar}\int \frac{d^3\vec{k}}{(2\pi)^3}\sum_n f_{n\vec{k}}\Omega^k_{n,ij}(\vec{k}),$$

$$\Omega^k_{n,ij}(\vec{k}) = -2Im\sum_{n\neq n'}\frac{\langle n\vec{k}|J^k_i|n'\vec{k}\rangle\langle n'\vec{k}|v_j|n\vec{k}\rangle}{\left(E_{n\vec{k}} - E_{n'\vec{k}}\right)^2},$$

where $f_{n\vec{k}}$ is the Fermi-Dirac distribution for the *n*th band, $J_i^k = \frac{1}{2}\{v_i, s_k\}$ is the spin current operator with spin operator $s_k$, $v_j = \frac{1}{\hbar}\frac{\partial H}{\partial k_j}$ is the velocity operator, and $i, j, k = x, y, z$. $\Omega_{n,ij}^k(\vec{k})$ is referred to as the spin Berry curvature in analogy to the ordinary Berry curvature. In order to calculate the spin Hall conductivities, we construct the tight-binding Hamiltonians using PAOFLOW code[23,24] based on the projection of the pseudo-atomic orbitals (PAO) from the non-self-consistent calculations with a 16 × 16 × 16 *k*-point mesh. The spin Hall conductivities were calculated using the tight-binding Hamiltonians with a 48 × 48 × 48 *k*-point mesh by the adaptive broadening method to get the converged values.

**Acknowledgments**

CBE acknowledges support for this research through a Vannevar Bush Faculty Fellowship (ONR N00014-20-1-2844), the Gordon and Betty Moore Foundation's EPiQS Initiative, Grant GBMF9065. Transport measurement at the University of Wisconsin–Madison was supported by the US Department of Energy (DOE), Office of Science, Office of Basic Energy Sciences (BES), under award number DE-FG02-06ER46327. Measurements at Cornell were supported by the DOE under award number DE-SC0017671. The work at UNL was partly supported by the National Science Foundation through the EPSCoR RII Track-1 program (Grant OIA-2044049). STEM measurement at Pohang University of Science and Technology was supported by the Basic Science Research Program through the National Research Foundation of Korea (NRF) funded by the Ministry of Science and ICT (2020R1A4A1018935).

**Figures and Tables**

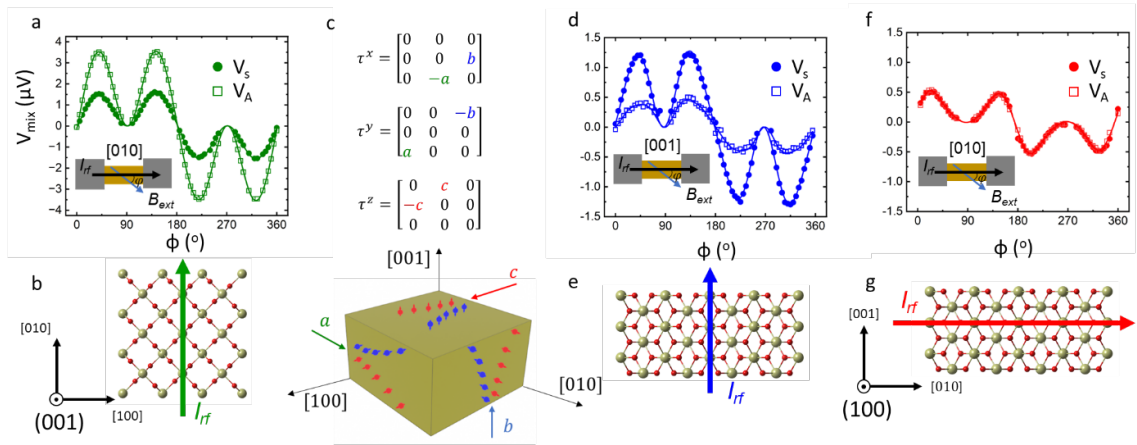

**Figure 1.** a, Symmetric and antisymmetric ST-FMR amplitudes as a function of magnetic-field angle for the (001) orientation with charge current along the [010] direction as depicted in b. c, experimental determination of the 3 non-zero STC elements using the (001) and (100) orientations. d, ST-FMR amplitudes as a function of magnetic-field angle for the (100) orientation with charge current along the [001] direction, as depicted in e. f, ST-FMR amplitudes as a function of magnetic-field angle for the (100) orientation with charge current along the [010] direction, as depicted in g.

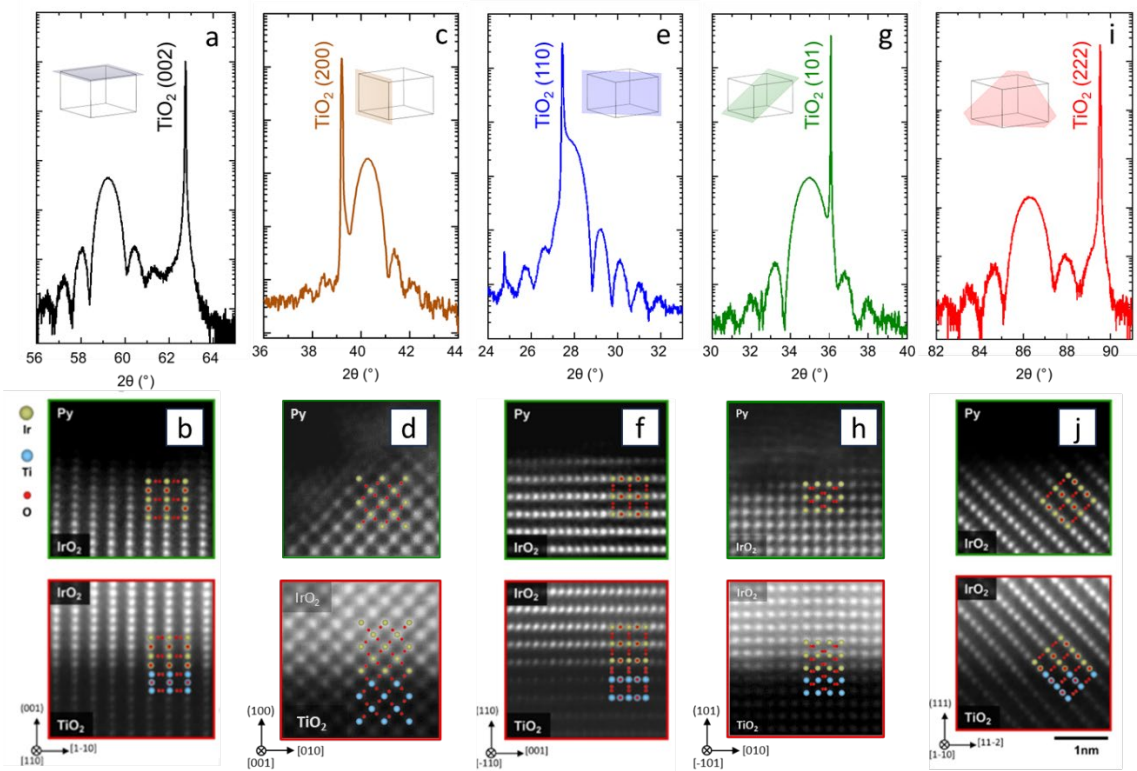

**Figure 2.** a, HR-XRD of the out of plane (002) peak and b, STEM of the interface between IrO$_2$ and Py and TiO$_2$ and IrO$_2$ with [1-10] zone axis. c, HR-XRD of the out of plane (200) peak and d, STEM of the interface between IrO2 and Py and TiO$_2$ and IrO$_2$ with [010] zone axis. e, HR-XRD of the out of plane (110) peak and f, STEM of the interface between IrO$_2$ and Py and TiO$_2$ and IrO$_2$ with [001] zone axis. g, HR-XRD of the out of plane (101) peak and h, STEM of the interface between IrO$_2$ and Py and TiO$_2$ and IrO$_2$ with [010] zone axis. And i, HR-XRD of the out of plane (222) peak and j, STEM of the interface between IrO$_2$ and Py and TiO$_2$ and IrO$_2$ with [11-2] zone axis.

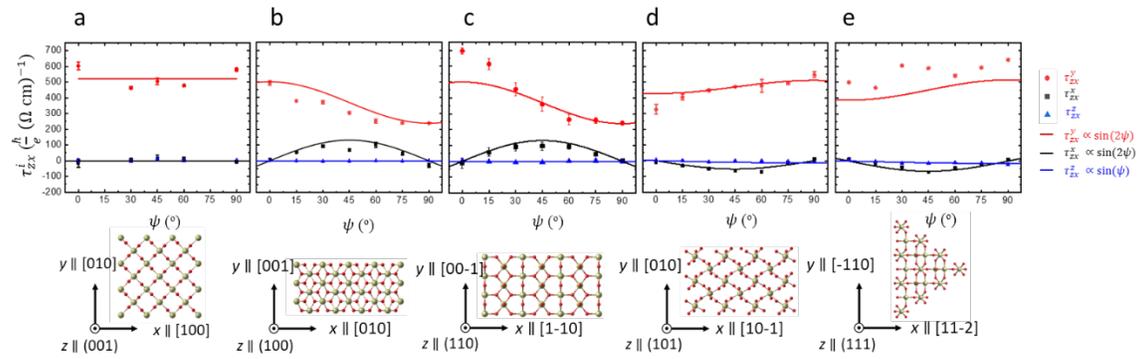

**Figure 3**: a-e, (001), (100), (110), (101), and (111) conventional and unconventional spin torque conductivities as a function of the angle $\psi$ of the in-plane electric field with respect to the x axis shown for each projection. Solid lines represent predictions based on rotating the experimental STC tensor, using the tensor elements determined from the ST-FMR results in the (001) and (100) orientations and the bulk symmetries of $IrO_2$.

# Supporting Information for

## Experimental tests of the full spin torque conductivity tensor in epitaxial $IrO_2$ thin films

Michael Patton, Daniel A. Pharis, Gautam Gurung, Xiaoxi Huang, Gahee Noh, Evgeny Y. Tsymbal, Si-Young Choi, Daniel C. Ralph, Mark S. Rzchowski, Chang-Beom Eom[*]

*Corresponding author: Chang-Beom Eom
Email: eom@engr.wisc.edu

**This PDF file includes:**

Supporting text
Figures S1 to S10
Table S1
SI References

**Supporting Information Text**

**Supplemental Note 1: Thin film growth and characterizations**
Epitaxial $IrO_2$ thin films 7-9 nm thick were grown on $TiO_2$ (001), (100), (110), (101), and (111) single crystal substrates by RF magnetron sputtering. The $IrO_2$ films were grown at 400°C at a pressure of 30 mTorr with 10% oxygen and a base pressure of $3 \times 10^{-7}$ Torr. The RF power was 20W. After growth the sample was cooled in an $O_2$ atmosphere. Ferromagnetic permalloy $Ni_{81}Fe_{19}$ (Py) 6-8 nm thick was then grown *in situ* at room temperature with 4 mTorr of Ar and an RF power of 35W. The devices were fabricated using photolithography and ion beam milling, followed by sputter deposition of 100 nm Pt/10nm Ti and lift off techniques for the electrodes.

X-ray diffraction reciprocal space mapping (RSM) was performed on $IrO_2$ (7-10nm) thin films for all five orientations as shown in Fig. S1. The (001) orientated $IrO_2$ shows coherent growth with slight relaxation (not shown). Fig. S1a and S1b show the (001) in-plane strain for the [010] and [001] direction, respectively, where we see fully coherent along the [010] and full relaxation along the [001] direction. Fig. S1c and S1d show the (110) in-plane strain for the [1-10] and [001] direction, respectively, where we see relaxation along both directions. Fig. S1e and S1f show the (101) in-plane strain for the [010] and [-101] direction, respectively, where we see fully coherent growth. Fig. S1g and S1h show the (111) in-plane strain along the [1-10] and [11-2] directions, respectively, where we see fully coherent growth.

Scanning transmission electron microscopy (STEM) was done for each orientation which demonstrates the sharp interface between $IrO_2$ and Py and between $IrO_2$ and $TiO_2$, which helps rule out any extrinsic effects that may contribute to the spin torques. Py/$IrO_2$/$TiO_2$ interfaces were visualized using STEM (JEM-ARM200F, JEOL) at 200 kV equipped with the aberration corrector (ASCOR, CEOS GmbH). The optimum size of the electron probe was set to be ~0.7 Å. The collection semi-angle of the HAADF detector was ranged from 54 to 216 mrad for clear Z-contrast images. The images were obtained using Smart Align and were conducted on multi-stacking images and aligned these images using rigid registration to correct for drift and scan distortions. The raw STEM images were filtered to reduce background noise by using Difference Filter (Filters Pro, HREM Research Inc., Japan). STEM samples were prepared by mechanical flat polishing and ion milling process. The polished samples were milled using a 3keV Ar ion beam. To minimize surface damage, the samples were milled with an acceleration voltage of 100 meV (PIPS II; Gatan, Pleasanton, CA, USA).

The crystal systems of rutile $IrO_2$ and rutile $TiO_2$ are the same tetragonal structure but different lattice constants (a=b=4.594 Å, c=2.959 Å for $TiO_2$; a=b=4.498 Å, c=3.154 Å for $IrO_2$) and thus the different strains could be accumulated depending on the substrate orientation. We analyze the strain states of $IrO_2$ films by using geometric phase analysis (GPA). In the case of $IrO_2$ (001) film, a and b lattice parameters are same and therefore the strain only through [110] projection is extracted, as shown in Fig. S2. The strain of IrO$_2$ film (Fig. S2b) is measured compared to the lattice parameter of TiO$_2$ and therefore the similar colors in IrO$_2$ and TiO$_2$ in Fig. S2b indicate that IrO$_2$ film is roughly matched with TiO$_2$ substrate. Thus, it should be noted that the measured strain from IrO$_2$ film (Fig. S2b) is not the real strain. To understand the actual strain, the lattice mismatch between the bulk IrO$_2$ and the substrate TiO$_2$ must be plotted together, as indicated by the green line in Fig. S2c. The black line, obtained by profiling the intensity from the rectangle box of Fig. S2b becomes comparable to bulk IrO$_2$ lattice parameter, which means that in-plane tensile strain along [1-10] is applied with slight relaxation.

IrO$_2$ (100) film experiences anisotropic strain states as shown in Fig. S3c and Fig. S3d because the lattice spacings along [001] and [010] are different to ones of TiO$_2$. The IrO$_2$ film along [001] is not fully relaxed, exhibiting ~ 2% compressive strain as shown in Fig. S3e, similar to the strain states in the (110) film along the [001] direction while the IrO$_2$ film along [010] is mostly coherent with the TiO$_2$ substrate as shown in Fig. S3f. IrO$_2$ (110) film is not coherently matched with the TiO$_2$ substrate as shown in Fig. S4c and Fig. S4d because the lattice spacings along [001] and [1-10] are different to ones of TiO$_2$. However, the IrO$_2$ film along [001] is not fully relaxed, exhibiting ~ 2.5% compressive strain as shown in Fig. S4e while the IrO$_2$ film along [1-10] is mostly relaxed as shown in Fig. S4f. IrO$_2$ (101) film is coherently matched with the TiO$_2$ substrate as shown in Fig. S5c and Fig. S5b. (101) appears to grow similarly with the (111) orientation which is the highest quality of all the orientations. Finally, for IrO$_2$ (111) film, the lattice spacing of IrO$_2$ along [11-2] is coherently matched with TiO$_2$ substrate as shown in Fig. S6c and Fig. S6e and therefore ~2% compressive strain along [11-2] exist in IrO$_2$ film. On the other hand, the lattice spacing of IrO$_2$ along [-110] is also mostly matched with that of the substrate, exhibiting ~1.8% tensile strain although this tensile strain is locally relaxed with an introduction of edge dislocations as indicated by the arrows in Fig. S6b and Fig. S6d.

**Supplemental Note 2: ST-FMR lineshape analysis**

The lineshape for IrO$_2$ (111) can be seen in Fig. S7a. The effective magnetization of Py can be obtained using Kittel's equation $f = \frac{\gamma}{2\pi}\sqrt{(H_{FMR} + H_K)(H_{FMR} + H_K + M_{eff})}$ where $\gamma$ is the gyromagnetic ratio and $H_K$ is the in-plane anisotropy field. We found the effective magnetization for (111) to be 0.85 T extracted from the fittings seen in Fig. S7b. The Gilbert damping coefficient $\alpha$ is obtained by fitting the linear relationship seen in Fig. S7c between the linewidth ($w$) and the frequency $w = w_0 + \left(\frac{2\pi}{\gamma}\right) * f$. To calibrate the microwave current ($I_{rf}$) we compare the resistance change due to Joule heating across the Hall bar device while varying the microwave power[123]. We can then find $I_{rf} = \sqrt{2}I_{dc}$ due to the joule heating relationship between ac and dc current. The anisotropic magnetoresistance (AMR) can be determined by measuring the device resistance as a function of in-plane magnetic field angle with applied magnetic field of 0.1T.

**Supplemental Note 3: Unconventional damping-like and field-like torques**

Zoomed in plots of the unconventional SOTs from the main text Fig. 3 for the (100), (110), (101) and (111) orientations can be seen in Fig. S8. Additional field-like x-SOTs were observed in the (100), (101), (110), and (111) as seen in Fig. S9 with Dresselhaus-like symmetry. We attribute this contribution to anisotropic resistivity within the material similar to other anisotropic materials such as 2D transition-metal dichalcogenides[4,5] where current transverse to the applied current in the device can arise due to anisotropic resistance. These currents then result in Oersted fields that can be analyzed in the angular ST-FMR as having $\sim \sin(2\varphi)\sin(\varphi)$ angular dependance in the $V_{A,mix}$ signal.

**Supplemental Note 4: DC-tuned ST-FMR**

DC tuned measurements were also performed on the (001) and (100) orientations for further validation of the SHC tensor. Some oxide substrates including $SrTiO_3$, $KTaO_3$, and $TiO_2$ have high dielectric loss resulting in current shunting when using high frequency rf currents, which can lead to a large symmetric $V_{mix}$ during ST-FMR measurements. DC tuned measurements have been used as an independent way to measure the spin Hall conductivity which applies a constant dc current in addition to the rf current during ST-FMR measurements, which can be used to isolate the true SHC of a material as the rf current shunting should not be affected by the applied dc current[6,7,8]. The linear relationship between the linewidth of the mixing voltage signal during ST-FMR measurements and the DC current can be used to determine the SHC. During the DC-tuned measurements, a DC bias was applied at currents ranging between -2 to 2 mA in addition to a fixed rf current by using a bias tee. The spin Hall angle can be determined using the following equation:

$$\theta_{DL} = \frac{2e}{\hbar}\left(\frac{(H_{FMR} + \frac{M_{eff}}{2})\mu_0 M_s t_{Py}}{\sin(\varphi)}\right)\frac{\Delta\alpha_{eff}}{\Delta J_c}$$

Where $t_{Py}$ is the thickness of permalloy and $\frac{\Delta\alpha_{eff}}{\Delta J_c}$ is the linear slope of effective damping coefficient, determined from the linear relationship between the linewidth ($w$) and the frequency $w = w_0 + \left(\frac{2\pi}{\gamma}\right)*f$, vs the charge current going through the $IrO_2$ layer determined using parallel resistor model. Fig. S10 shows the DC tuned results for (001) oriented $IrO_2$, (100) oriented $IrO_2$ with current along the [001] direction, and (100) oriented $IrO_2$ with current along the [010] direction and the results are summarized in Table 1. Although the DC-tuned results are in good agreement with the theoretical calculations, we chose to use the angular ST-FMR results for the experimental SHC tensor to compare to the other orientations in the main text as determining the unconventional SHCs can be difficult and not straight forward using DC-tuned techniques.

## Supplemental Note 5: Tensor rotation from the (001) basis to other orientations

Spin Torque Conductivity for IrO$_2$ (001)
$\hat{X} = [100], \hat{Y} = [010], \hat{Z} = [001]$

$$\tau^x = \begin{pmatrix} 0 & 0 & 0 \\ 0 & 0 & b \\ 0 & -a & 0 \end{pmatrix} \quad \tau^y = \begin{pmatrix} 0 & 0 & -b \\ 0 & 0 & 0 \\ a & 0 & 0 \end{pmatrix} \quad \tau^z = \begin{pmatrix} 0 & c & 0 \\ -c & 0 & 0 \\ 0 & 0 & 0 \end{pmatrix}$$

For $E \parallel \cos\psi\hat{x} + \sin\psi\hat{y}$, the in-plane spin $= -a\sin\psi\hat{x} + a\cos\psi\hat{y}$.
$\tau_{zx}^y = (-\sin\psi\hat{x} + \cos\psi\hat{y}) \cdot (-a\sin\psi\hat{x} + a\cos\psi\hat{y}) = a$.
$\tau_{zx}^x = 0$.
$\tau_{zx}^z = 0$.

Spin Torque Conductivity for IrO$_2$ (100)
$\hat{X} = [010], \hat{Y} = [001], \hat{Z} = [100]$

$$R = \begin{pmatrix} 0 & 1 & 0 \\ 0 & 0 & 1 \\ 1 & 0 & 0 \end{pmatrix}$$

$$\tau^x = \begin{pmatrix} 0 & 0 & 0 \\ 0 & 0 & a \\ 0 & -b & 0 \end{pmatrix} \quad \tau^y = \begin{pmatrix} 0 & 0 & -c \\ 0 & 0 & 0 \\ c & 0 & 0 \end{pmatrix} \quad \tau^z = \begin{pmatrix} 0 & b & 0 \\ -a & 0 & 0 \\ 0 & 0 & 0 \end{pmatrix}$$

For $E \parallel \cos\psi\hat{x} + \sin\psi\hat{y}$, the in-plane spin $= -b\sin\psi\hat{x} + c\cos\psi\hat{y}$.
$\tau_{zx}^y = (-\sin\psi\hat{x} + \cos\psi\hat{y}) \cdot (-b\sin\psi\hat{x} + c\cos\psi\hat{y}) = (b\sin^2\psi + c\cos^2\psi) = b + \cos^2\psi(c - b)$
$\quad = 0.5(b + c) + 0.5(c - b)\cos 2\psi$
$\tau_{zx}^x = (\cos\psi\hat{x} + \sin\psi\hat{y}) \cdot (-b\sin\psi\hat{x} + c\cos\psi\hat{y}) = \sin\psi\cos\psi(c - b) = 0.5(c - b)\sin 2\psi$.
$\tau_{zx}^z = 0$.

Spin Torque Conductivity for IrO$_2$ (101)
$\hat{X} = \cos\theta[100] - \sin\theta[001], \hat{Y} = [010], \hat{Z} = \sin\theta[100] + \cos\theta[001]$
$\tan\theta = [001]$ lattice spacing/$[100]$ lattice spacing $= 0.7012, \sin\theta = 0.574, \cos\theta = 0.819$

$$R = \begin{pmatrix} \cos\theta & 0 & -\sin\theta \\ 0 & 1 & 0 \\ \sin\theta & 0 & \cos\theta \end{pmatrix}$$

$$\tau^x = \begin{pmatrix} 0 & -\sin\theta\cos\theta(c - a) & 0 \\ \sin\theta\cos\theta(c - b) & 0 & b\cos^2\theta + c\sin^2\theta \\ 0 & -a\cos^2\theta - c\sin^2\theta & 0 \end{pmatrix}$$

$$\tau^y = \begin{pmatrix} -\sin\theta\cos\theta(a - b) & 0 & -b\cos^2\theta - a\sin^2\theta \\ 0 & 0 & 0 \\ b\sin^2\theta + a\cos^2\theta & 0 & \sin\theta\cos\theta(a - b) \end{pmatrix}$$

$$\tau^z = \begin{pmatrix} 0 & a\sin^2\theta + c\cos^2\theta & 0 \\ -b\sin^2\theta - c\cos^2\theta & 0 & -\sin\theta\cos\theta(c - b) \\ 0 & -\sin\theta\cos\theta(a - c) & 0 \end{pmatrix}$$

For $E \parallel \cos\psi\hat{x} + \sin\psi\hat{y}$, the in-plane spin $= \sin\psi(-a\cos^2\theta - c\sin^2\theta)\hat{x} + \cos\psi(b\sin^2\theta + a\cos^2\theta)\hat{y}$.

$$\tau_{zx}^{y} = (-\sin\psi\hat{x} + \cos\psi\hat{y}) \cdot (\sin\psi(-a\cos^2\theta - c\sin^2\theta)\hat{x} + \cos\psi(b\sin^2\theta + a\cos^2\theta)\hat{y})$$
$$= (a\cos^2\theta + c\sin^2\theta)\sin^2\psi + (b\sin^2\theta + a\cos^2\theta)\cos^2\psi$$
$$= (a\cos^2\theta + c\sin^2\theta) + \cos^2\psi\sin^2\theta(b - c)$$
$$= (a\cos^2\theta + c\sin^2\theta) + (1 + \cos2\psi)\sin^2\theta(b - c)/2$$
$$= \left(a\cos^2\theta + \frac{(c+b)}{2}\sin^2\theta\right) + \cos2\psi\sin^2\theta(b - c)/2$$
$$= 0.670a + 0.165(c + b) - \cos2\psi\, 0.165(c - b)$$
$$\tau_{zx}^{x} = (\cos\psi\hat{x} + \sin\psi\hat{y}) \cdot (\sin\psi(-a\cos^2\theta - c\sin^2\theta)\hat{x} + \cos\psi(b\sin^2\theta + a\cos^2\theta)\hat{y})$$
$$= -\sin\psi\cos\psi\sin^2\theta(c - b).$$
$$= -\sin2\psi\, 0.165(c - b).$$
$$\tau_{zx}^{z} = -\sin\psi\, \sin\theta\cos\theta(a - c) = -\sin\psi\, 0.47(a - c).$$

Spin Torque Conductivity for IrO$_2$ (110)

$\hat{X} = \frac{1}{\sqrt{2}}([100] - [010]), \quad \hat{Y} = [00\bar{1}], \quad \hat{Z} = \frac{1}{\sqrt{2}}([100] + [010])$

$$R = \begin{pmatrix} \frac{1}{\sqrt{2}} & -\frac{1}{\sqrt{2}} & 0 \\ 0 & 0 & -1 \\ \frac{1}{\sqrt{2}} & \frac{1}{\sqrt{2}} & 0 \end{pmatrix}$$

$$\tau^{x} = \begin{pmatrix} 0 & 0 & 0 \\ 0 & 0 & a \\ 0 & -b & 0 \end{pmatrix} \quad \tau^{y} = \begin{pmatrix} 0 & 0 & -c \\ 0 & 0 & 0 \\ c & 0 & 0 \end{pmatrix} \quad \tau^{z} = \begin{pmatrix} 0 & b & 0 \\ -a & 0 & 0 \\ 0 & 0 & 0 \end{pmatrix}$$

For E ∥ $\cos\psi\hat{x} + \sin\psi\hat{y}$, the in-plane spin = $-\sin\psi b\hat{x} + \cos\psi c\hat{y}$.

$\tau_{zx}^{y} = (-\sin\psi\hat{x} + \cos\psi\hat{y}) \cdot (-\sin\psi b\hat{x} + \cos\psi c\hat{y}) = b\sin^2\psi + c\cos^2\psi = b + \cos^2\psi(c - b)$
$\quad = 0.5(b + c) + \cos2\psi\, 0.5(c - b)$
$\tau_{zx}^{x} = (\cos\psi\hat{x} + \sin\psi\hat{y}) \cdot (-\sin\psi b\hat{x} + \cos\psi c\hat{y}) = 0.5\sin2\psi(c - b).$
$\tau_{zx}^{z} = 0.$

Spin Torque Conductivity for IrO$_2$ (111)

$R \equiv$ [100] lattice spacing/[001] lattice spacing $= 1.426$.

$\hat{Z} = \frac{1}{\sqrt{2+(R)^2}}([100] + [010] + (R)[001]) = 0.498[100] + 0.498[010] + 0.710[001]$

$\hat{Y} = -0.707[100] + 0.707[010] + 0[001]$

$\hat{X} = [100] + [010] - 0.996(0.498[100] + 0.498[010] + 0.710[001])$
$\quad\quad\quad \to 0.502[100] + 0.502[010] - 0.704[001]$

$$R = \begin{pmatrix} 0.502 & 0.502 & -0.704 \\ -0.707 & +0.707 & 0 \\ 0.498 & 0.498 & 0.710 \end{pmatrix}$$

$$\tau^{x} = 0.502\left[\begin{pmatrix} 0.3535(a - b) & 0.498a & 0.3565b + 0.351a \\ -0.498b & 0 & 0.502b \\ -0.351b - 0.3565a & -0.502a & -0.3535(a - b) \end{pmatrix}\right]$$
$$+ 0.502\left[\begin{pmatrix} -0.3535(a - b) & 0.498a & -0.3565b - 0.351a \\ -0.498b & 0 & 0.502b \\ 0.351b + 0.3565a & -0.502a & 0.3535(a - b) \end{pmatrix}\right]$$

$$-0.704\left[\begin{pmatrix} 0 & 0.710c & 0 \\ -0.710c & 0 & -0.704c \\ 0 & 0.704c & 0 \end{pmatrix}\right]$$

$$= \left[\begin{pmatrix} 0 & 0.5(a-c) & 0 \\ 0.5(c-b) & 0 & 0.504b + 0.496c \\ 0 & -0.504a - 0.496c & 0 \end{pmatrix}\right]$$

$$\tau^y_{\square} = -0.707\left[\begin{pmatrix} 0.3535(a-b) & 0.498a & 0.3565b + 0.351a \\ -0.498b & 0 & 0.502b \\ -0.351b - 0.3565a & -0.502a & -0.3535(a-b) \end{pmatrix}\right]$$

$$+0.707\left[\begin{pmatrix} -0.3535(a-b) & 0.498a & -0.3565b - 0.351a \\ -0.498b & 0 & 0.502b \\ 0.351b + 0.3565a & -0.502a & 0.3535(a-b) \end{pmatrix}\right]$$

$$= \left[\begin{pmatrix} -0.5(a-b) & 0 & -0.504b - 0.496a \\ 0 & 0 & 0 \\ 0.504a + 0.496b & 0 & 0.5(a-b) \end{pmatrix}\right]$$

$$\sigma^z = 0.498\left[\begin{pmatrix} 0.3535(a-b) & 0.498a & 0.3565b + 0.351a \\ -0.498b & 0 & 0.502b \\ -0.351b - 0.3565a & -0.502a & -0.3535(a-b) \end{pmatrix}\right]$$

$$+0.498\left[\begin{pmatrix} -0.3535(a-b) & 0.498a & -0.3565b - 0.351a \\ -0.498b & 0 & 0.502b \\ 0.351b + 0.3565a & -0.502a & 0.3535(a-b) \end{pmatrix}\right]$$

$$+0.710\left[\begin{pmatrix} 0 & 0.710c & 0 \\ -0.710c & 0 & -0.704c \\ 0 & 0.704c & 0 \end{pmatrix}\right]$$

$$= \left[\begin{pmatrix} 0 & 0.496a + 0.504c & 0 \\ -0.496b - 0.504c & 0 & -0.5(c-b) \\ 0 & -0.5(a-c) & 0 \end{pmatrix}\right]$$

For E ∥ $\cos\psi\hat{x} + \sin\psi\hat{y}$, the in-plane spin $=[\sin\psi(-0.504a - 0.496c)]\hat{x} + [\cos\psi(0.504a + 0.496b)]\hat{y}$.

$\tau^y_{zx} = (-\sin\psi\hat{x} + \cos\psi\hat{y})$
$\qquad \cdot ([\sin\psi(-0.504a - 0.496c)]\hat{x} + [\cos\psi(0.504a + 0.496b)]\hat{y})$
$\quad = \sin^2\psi(0.504a + 0.496c) + \cos^2\psi(0.504a + 0.496b)$
$\quad = (0.504a + 0.496c) - \cos^2\psi(0.496(c-b))$
$\quad = 0.504a + 0.248c + 0.248b - \cos 2\psi(0.248(c-b))$

$\tau^x_{zx} = (\cos\psi\hat{x} + \sin\psi\hat{y}) \cdot ([\sin\psi(-0.504a - 0.496c)]\hat{x} + [\cos\psi(0.504a + 0.496b)]\hat{y}) =$
$\quad = -\cos\psi\sin\psi\, 0.496(c-b) = -\sin 2\psi\, 0.248(c-b)$

$\tau^z_{zx} = -\sin\psi(0.5(a-c))$

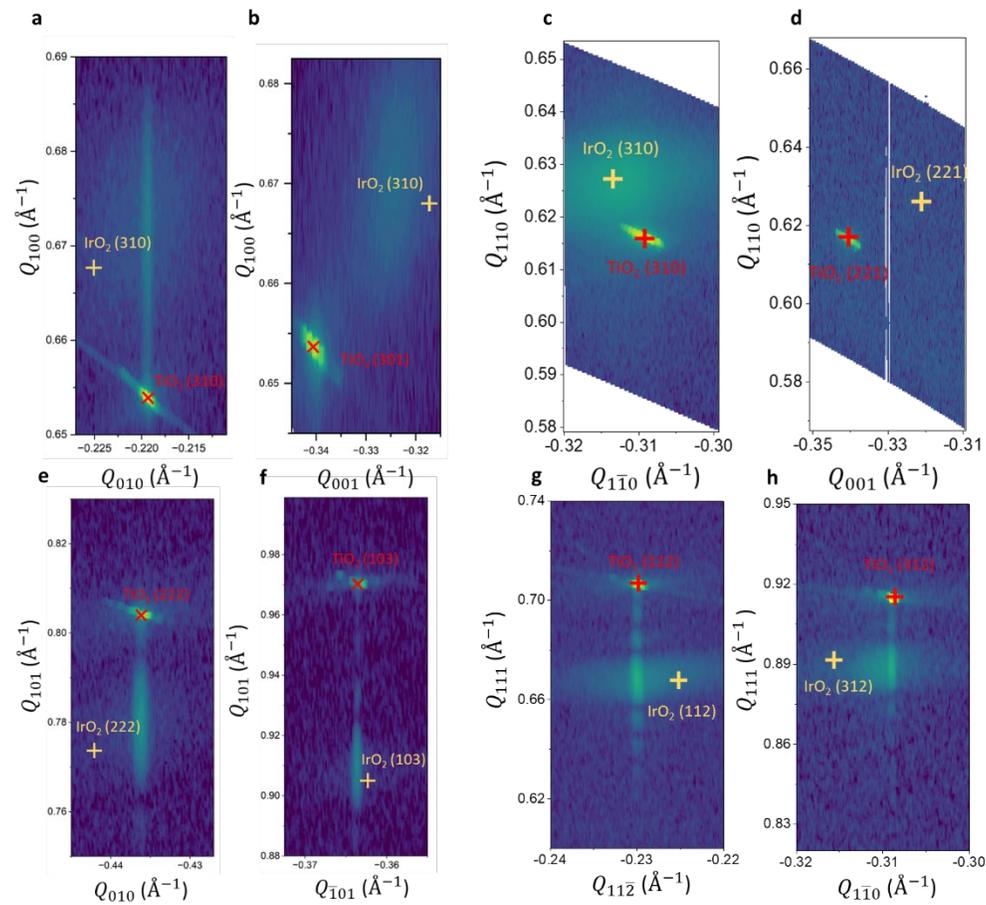

Figure S1: a and b RSMs of the (100) orientation, c and d RSMs of the (110) orientation, e and f RSMs of the (101) orientation, g and h and g RSMs of the (111) orientation.

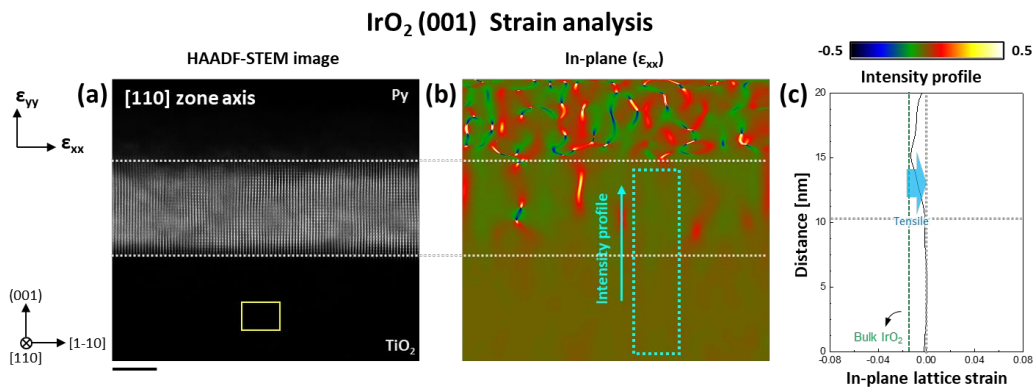

Figure S2: Strain analysis in IrO$_2$ (001) film. a, HAADF-STEM image of IrO$_2$ (001) films in [110] projection. Scale bar is 5nm. b, Map of in-plane ($\varepsilon_{xx}$) strain analysis of a. c, Intensity profile of In-plane (GPA) image b. The lattice strain is calculated based on the lattice parameter of reference region, which is yellow box in TiO$_2$ substrate. The interfaces between IrO$_2$/TiO$_2$ and Py/IrO$_2$ are identified by white dashed line. In-plane strain in [1-10] direction is roughly applied.

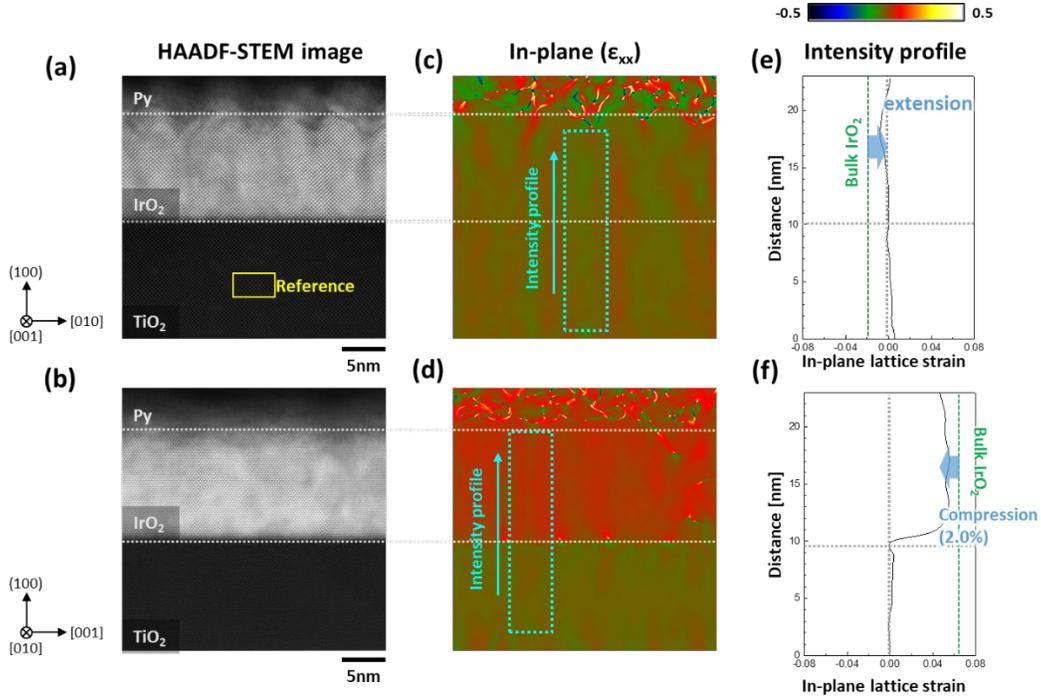

Figure S3: Strain analysis in IrO$_2$ (100) film. a, HAADF-STEM image of IrO$_2$ (100) films in [110] projection. Scale bar is 5nm. b, Map of in-plane ($\varepsilon_{xx}$) strain analysis of a. c, Intensity profile of In-plane (GPA) image b. The lattice strain is calculated based on the lattice parameter of reference region, which is yellow box in TiO$_2$ substrate. The interfaces between IrO$_2$/TiO$_2$ and Py/IrO$_2$ are identified by white dashed line.

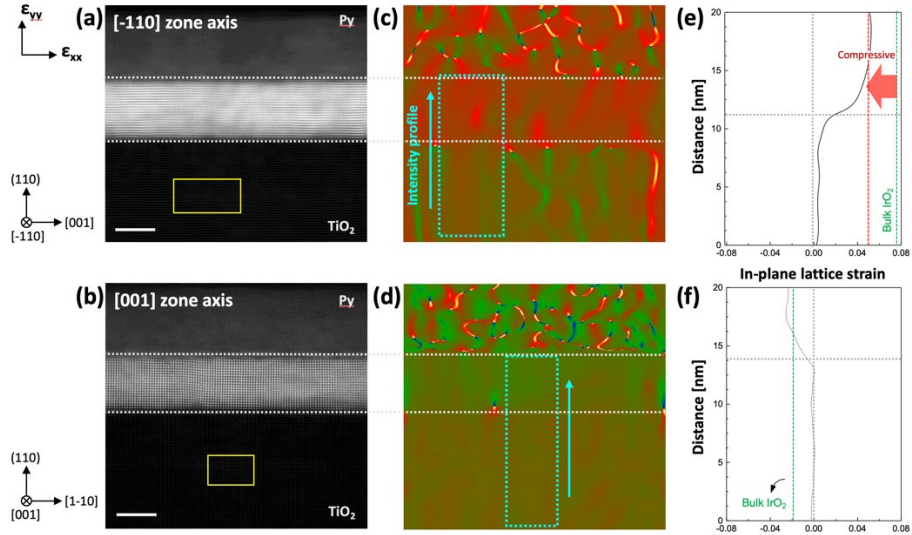

Figure S4: Strain analysis in IrO$_2$ (110) film. a, b HAADF-STEM image of IrO$_2$ (110) films in [001] and [-110] projection. Scale bar is 5nm. c, d Map of in-plane ($\varepsilon_{xx}$) strain analysis of a and b. e, f Intensity profile of In-plane GPA image c and d. The lattice strain is calculated based on the lattice parameter of reference region, which is yellow box in TiO$_2$ substrate. The interfaces between IrO$_2$/TiO$_2$ and Py/IrO$_2$ are identified by white dashed line. Tensile strain in [1-10] direction is fully relaxed, but compressive strain is partially relaxed. A compressive strain of about 2.5% is applied in the [001] direction.

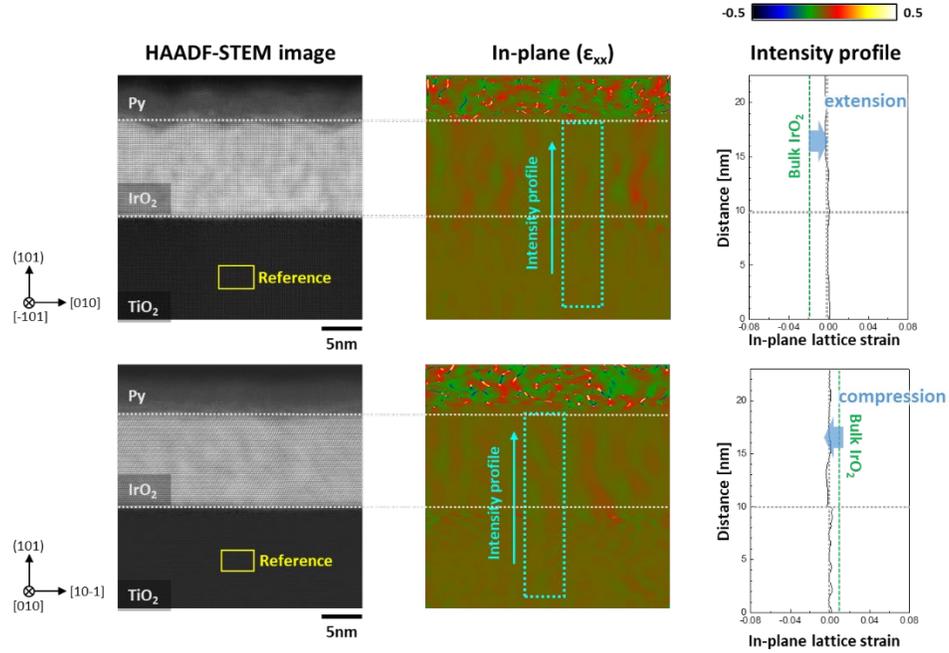

Figure S5: Strain analysis in IrO$_2$ (101) film. a, HAADF-STEM image of IrO$_2$ (101) films in [110] projection. Scale bar is 5nm. b, Map of in-plane ($\varepsilon_{xx}$) strain analysis of a. c, Intensity profile of In-plane (GPA) image b. The lattice strain is calculated based on the lattice parameter of reference region, which is yellow box in TiO$_2$ substrate. The interfaces between IrO$_2$/TiO$_2$ and Py/IrO$_2$ are identified by white dashed line.

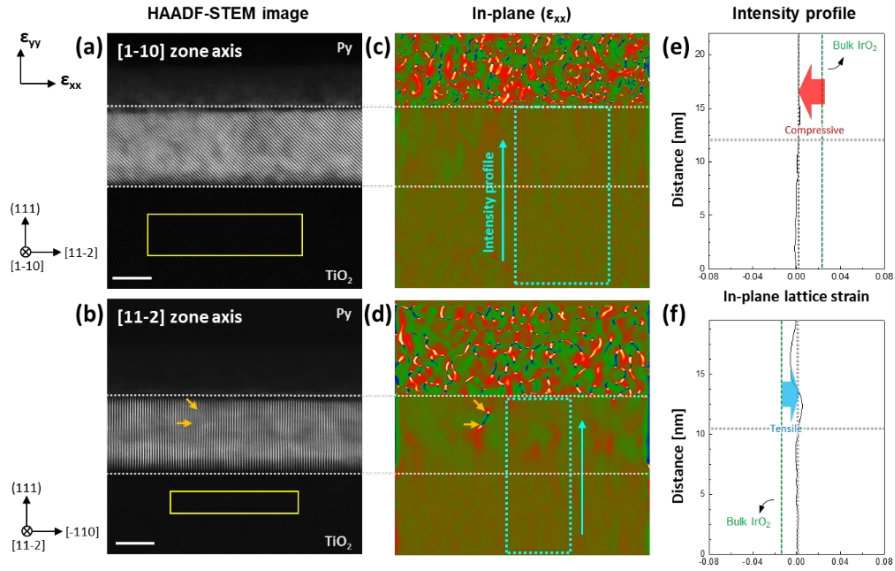

Figure S6: Strain analysis in IrO$_2$ (111) film. a, b HAADF-STEM image of IrO$_2$ (111) films in [1-10] and [11-2] projection. Scale bar is 5nm. c, d Map of in-plane ($\varepsilon_{xx}$) strain analysis of a and b. e, f Intensity profile of In-plane GPA image c and d. The lattice strain is calculated based on the lattice parameter of reference region, which is yellow box in TiO$_2$ substrate. The interfaces between IrO$_2$/TiO$_2$ and Py/IrO$_2$ are identified by white dashed line. Tensile strain in [1-10] direction and compressive strain in [11-2] direction are fully applied.

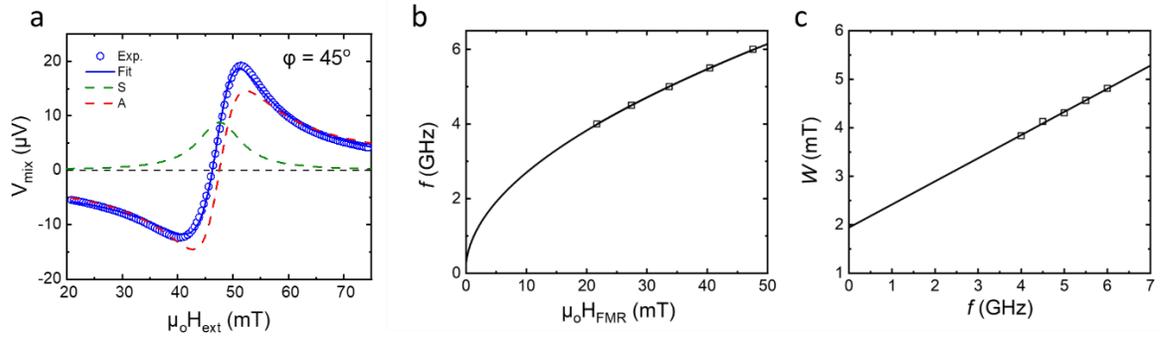

Figure S7: a, lineshape analysis for IrO$_2$(111)/Py showing the V$_s$ and V$_A$ from the ST-FMR measurement. b, Resonance field vs RF frequency with Kittel formula used for fitting. And c, linear relationship of the line width vs RF frequency to obtain the Gilbert damping coefficient.

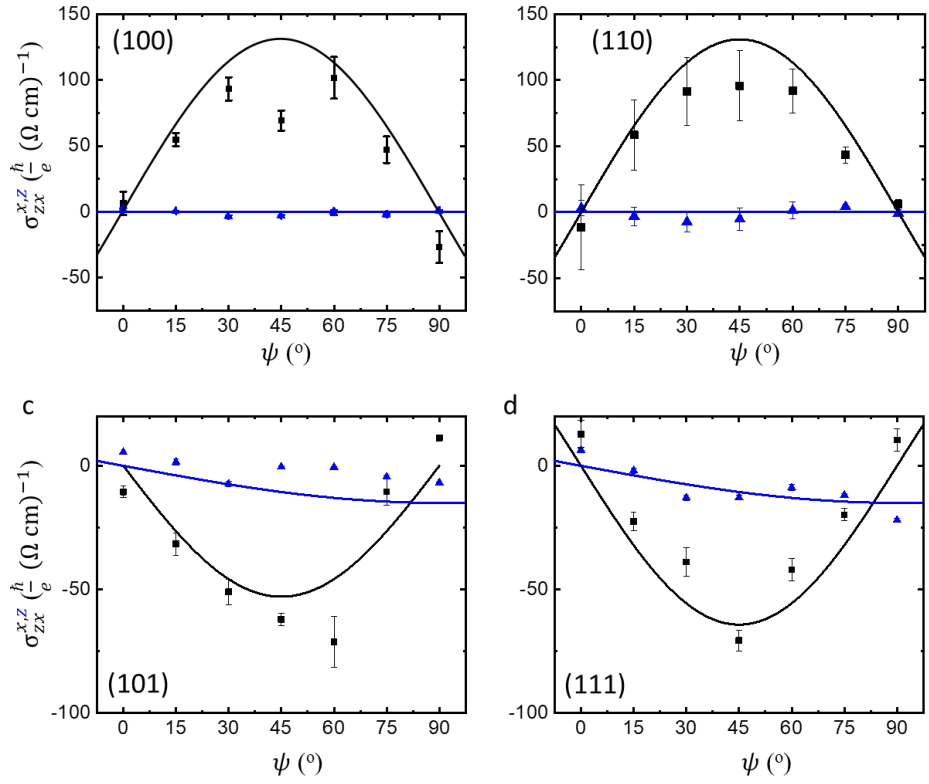

Figure S8: a-d, unconventional damping-like x-SOTs (black) and unconventional damping-like z-SOTs (blue) for (100), (110), (101), and (111), respectively vs in-plane direction of the charge current.

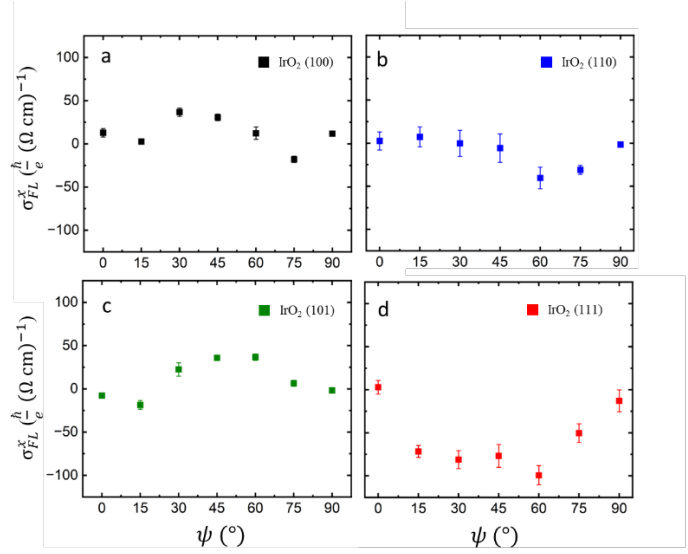

Figure S9: a-d, unconventional field-like x-SOTs for (100), (110), (101), and (111), respectively vs in-plane direction of the charge current.

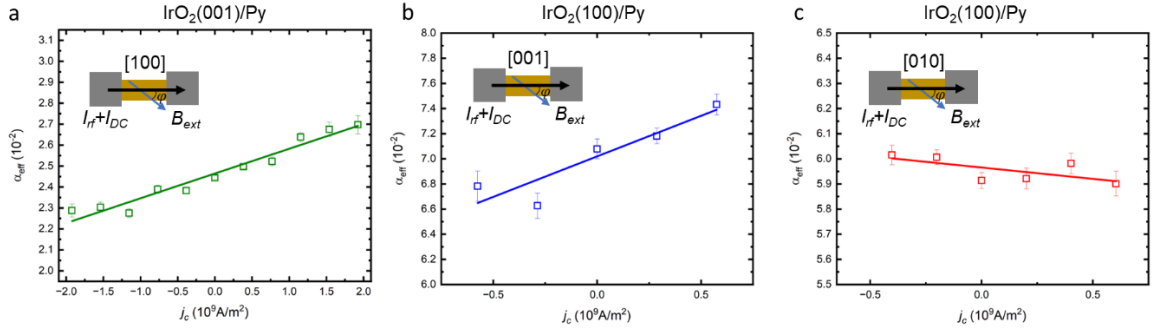

Figure S10: a, DC-tuned ST-FMR for the (001) orientation, b, (100) orientation with current along the [001] direction, and c, (100) orientation with current along the [010] direction where $\alpha_{eff}$ is the effective gilbert damping coefficient and $j_c$ is the DC current going through the $IrO_2$ layer.

Table S1: Summary of angular and DC-tune ST-FMR results for the (001) and (100) orientations giving the 3 non-zero SHC terms *a* ,*b* and *c*.

| Orientation | SHC term | Method | $\theta^y$ | $\sigma^y$ ($\frac{\hbar}{e}$ ($\Omega$ cm)$^{-1}$) |
|---|---|---|---|---|
| IrO$_2$(001) $j_c$ ∥ [010] | a | Angular ST-FMR | 0.12±0.01 | 520 ± 19 |
| IrO$_2$(001) $j_c$ ∥ [010] | a | DC-tuned | 0.035± 0.003 | 247 ± 21 |
| IrO$_2$(100) $j_c$ ∥ [001] | b | Angular ST-FMR | 0.08 ± 0.002 | 238 ± 5 |
| IrO$_2$(100) $j_c$ ∥ [001] | b | DC-tuned | 0.06 ± 0.015 | 190 ± 50 |
| IrO$_2$(100) $j_c$ ∥ [010] | c | Angular ST-FMR | 0.20±0.006 | 493 ± 15 |
| IrO$_2$(100) $j_c$ ∥ [010] | c | DC-tuned | -0.03 ± 0.006 | -34 ± 14 |